\documentclass[preprint,showpacs,preprintnumbers,amsmath,amssymb]{revtex4}
\usepackage{amssymb}
\usepackage{amsmath}
\usepackage{graphicx}
\usepackage{epsfig}
\usepackage{subfigure}
\usepackage{amsfonts}
\usepackage{CJK}

\begin{document}

\title{One-step implementation of a multi-target-qubit controlled phase gate with cat-state qubits in circuit QED}

\author{You-Ji Fan$^{1}$}
\author{Zhen-Fei Zheng$^{2}$}
\author{Yu Zhang$^{3}$}
\author{Dao-Ming Lu$^{1}$}
\author{Chui-Ping Yang$^{4,5}$}
\email{yangcp@hznu.edu.cn}

\address{$^1$College of Mechanic and Electronic Engineering, Wuyi University, Wuyishan 354300, Fujiang, China}
\address{$^2$CAS Key Laboratory of Quantum Information, University of Science and Technology of China, Hefei 230026, China}
\address{$^3$School of Physics, Nanjing University, Nanjing 210093, China}
\address{$^4$Quantum Information Research Center, Shangrao Normal University, Shangrao 334001, China}
\address{$^5$Department of Physics, Hangzhou Normal University, Hangzhou 310036, China}

\begin{abstract}
We propose a single-step implementation of a muti-target-qubit controlled phase gate with
one cat-state qubit (\textit{cqubit}) simultaneously controlling $n-1$ target
\textit{cqubits}. The two logic states of a \textit{cqubit} are represented
by two orthogonal cat states of a single cavity mode. In this proposal, the
gate is implemented with $n$ microwave cavities coupled to a
superconducting transmon qutrit. Because the qutrit remains in the ground
state during the gate operation, decoherence caused due to the qutrit's
energy relaxation and dephasing is greatly suppressed. The gate
implementation is quite simple because only a single-step operation is needed
and neither classical pulse nor measurement is required. Numerical
simulations demonstrate that high-fidelity realization of a controlled phase gate with
one cqubit simultaneously controlling two target cqubits is
feasible with present circuit QED technology. This proposal can be extended
to a wide range of physical systems to realize the proposed gate, such as
multiple microwave or optical cavities coupled to a natural or artificial
three-level atom.
\end{abstract}
\maketitle
\date{\today}

\section{Introduction}

Quantum computing has attracted considerable interest since quantum
computers can solve hard computational problems much more efficiently than
classical computers [1-3]. Multiqubit gates play an important role in
quantum computing. It is known that a multiqubit gate can in principle be
constructed with single-qubit and two-qubit basic gates. However, the
methods based on the conventional gate-decomposition protocols [4-6] are
complicated and not easy to implement experimentally. For instance, the
number of universal two-qubit gates, which are needed to implement
multiqubit gates, increases drastically with the number of qubits [4-6]. As
a result, the operation time required for implementing a multiqubit gate
would be quite long and thus the fidelity would be significantly
deteriorated by decoherence. Hence, it is worthwhile to seek efficient
approaches to realize multiqubit gates.

There exist two kinds of significant multiqubit gates, i.e., multiqubit
gates with multiple control qubits acting on a single target qubit (also
called multiqubit Toffoli gates or multi-control-qubit gates), and
multiqubit gates with a single qubit simultaneously controlling multiple
target qubits. For the past years, much progress has been made in the
physical realization of these two types of multiqubit gates. Several schemes
for realizing three-qubit Toffoli gates have been proposed with neutral
atoms in an optical lattice \cite{s7} or hybrid atom-photon qubits \cite{s8}%
. In addition, experimental realization of a three-qubit controlled phase
gate in NMR quantum system \cite{s9} and a three-qubit Toffoli gate with
superconducting qubits \cite{s10} has been reported. On the other hand,
based on cavity or circuit QED, many theoretical proposals have been
presented for directly realizing not only multi-control-qubit gates [11-22]
but also multi-target-qubit gates [23-27], in various physical qubits.

In recent years, cat-state qubits (\textit{cqubits}), which are encoded with
cat states, have drawn intensive attention due to their enhanced life times
with quantum error correction (QEC). For instance, the lifetime of a cqubit
can be made to be $2$ up to $320$ $\mu $s with QEC \cite{s28}. Recently,
there is an increasing interest in quantum computing with cat-state encoding
qubits. Several schemes have been presented for realizing single-cqubit
gates and two-cqubit gates [29-31]. Moreover, single-cqubit gates \cite{s32}
and two-cqubit entangled Bell states \cite{s33} have been experimentally
implemented recently. In addition, the circuit QED, consisting of microwave cavities and artificial atoms,
is particularly attractive and has been considered as one of the leading candidates for quantum
information processing [34-45].

The focus of this work is on a multi-target-qubit controlled phase gate with
one qubit simultaneously controlling multiple target qubits. This multi
qubit phase gate is described by

\begin{eqnarray}
\left\vert 0_{1}\right\rangle \left\vert i_{2}\right\rangle \left\vert
i_{3}\right\rangle ...\left\vert i_{n}\right\rangle &\rightarrow &\left\vert
0_{1}\right\rangle \left\vert i_{2}\right\rangle \left\vert
i_{3}\right\rangle ...\left\vert i_{n}\right\rangle ,  \nonumber \\
\left\vert 1_{1}\right\rangle \left\vert i_{2}\right\rangle \left\vert
i_{3}\right\rangle ...\left\vert i_{n}\right\rangle &\rightarrow &\left\vert
1_{1}\right\rangle \left( -1\right) ^{i_{2}}\left( -1\right)
^{i_{3}}...\left( -1\right) ^{i_{n}}\left\vert i_{2}\right\rangle \left\vert
i_{3}\right\rangle ...\left\vert i_{n}\right\rangle ,
\end{eqnarray}
where subscript $1$ represents the control qubit while subscripts $%
2,3,\ldots ,$ and $n$ represent target qubits, and $i_{2},i_{3},...i_{n}\in
\left\{ 0,1\right\} $. Equation (1) implies that, when the control qubit is
in the state $\left\vert 0\right\rangle $, nothing happens to the states of
each target qubit; however, when the control qubit is in $\left\vert
1\right\rangle $, a phase flip (from sign $+$ to $-$) happens to the state $%
\left\vert 1\right\rangle $ of each target qubit. This multiqubit gate is
useful in quantum computing and quantum information processing, such as in
entanglement preparation \cite{s46}, error correction \cite{s47}, quantum
algorithms \cite{s48}, and quantum cloning \cite{s49}. After a deep search
of the literature, we found that how to directly realize this multiqubit
gate with cat-state qubits has not been reported yet.

Motivated by the above, we will propose a method to realize the
multi-target-qubit controlled phase gate (1) with \textit{cqubits}, by using
$n$ microwave cavities coupled to a superconducting transmon qutrit (a
three-level artificial atom) (Fig.~1). This proposal is based on circuit
QED. As shown below, this proposal has the following advantages. During the
gate operation, the qutrit stays in the ground state. Thus, decoherence from
the qutrit is greatly suppressed. The gate implementation is simple because
of only one-step operation and no need of classical pulse or measurement.
The gate operation time is independent of the number of the \textit{cqubits}%
. Our numerical simulations show that high-fidelity implementation of a
controlled phase with one cqubit simultaneously controlling two cqubits is
feasible with current circuit QED technology. This proposal can be extended
to a wide range of physical systems to realize the proposed gate, such as
multiple microwave or optical cavities coupled to a natural or artificial
three-level atom.

This paper is organized as follows. In Sec. \ref{sec-II}, we explicitly show
how to realize a controlled-phase gate with one cqubit simultaneously
controlling $n-1$ target cqubits. In Sec. \ref{sec-III}, we give a brief
discussion on the experimental feasibility of implementing a three-qubit
controlled phase gate with one cqubit simultaneously controlling two target
cqubits. We end up with a conclusion in Sec. \ref{sec-con}.

\begin{figure}[tbp]
\begin{center}
\includegraphics[bb=38 391 532 608, width=12.5 cm, clip]{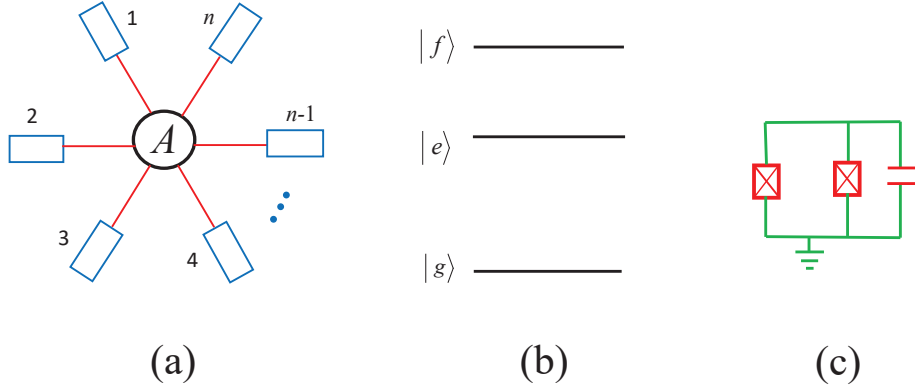} \vspace*{%
-0.08in}
\end{center}
\caption{(Color online) (a) Diagram of $n$ cavities ($1,2,...,n$) coupled to
a superconducting transmon qutrit (A). A square represents a cavity, which
can be one-dimensional or three-dimensional cavity. The qutrit is
capacitively or inductively coupled to each resonator. (b) Level
configuration of the transmon qutrit, for which the level spacing between
the upper two levels is smaller than that between the two lowest levels. (c)
Electronic circuit of a transmon qutrit, which consists of two Josephson
junctions and a capacitor.}
\label{fig1}
\end{figure}

\section{Multi-Target-cqubit Controlled Phase Gate}

\label{sec-II} 
\begin{figure}[tbp]
\begin{center}
\includegraphics[bb=152 560 372 733, width=8.0 cm, clip]{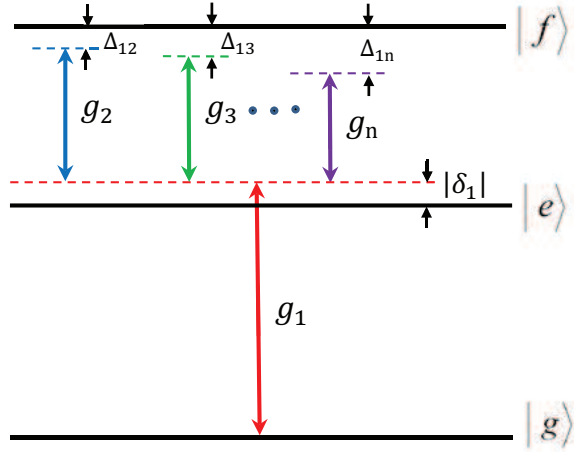} \vspace*{%
-0.08in}
\end{center}
\caption{(Color online) Cavity $1$ is far-off resonant with the $|g\rangle
\leftrightarrow |e\rangle $ transition of the qutrit with coupling strength $%
g_{1}$ and detuning $\left\vert \protect\delta _{1}\right\vert $, while
cavity $l$ ($l=2,3,...,n$) is far-off resonant with the $|e\rangle
\leftrightarrow |f\rangle $ transition of the qutrit with coupling strength $%
g_{l}$ and detuning $\left\vert \protect\delta _{l}\right\vert $. Note that
the detunings $\left\vert \protect\delta _{l}\right\vert $ ($l=2,3,...,n$)
are not drawn to simplify the figure. From the figure, one can see $%
\left\vert \protect\delta _{1}\right\vert =\protect\omega _{c_{1}}-\protect%
\omega _{eg}$, $\left\vert \protect\delta _{l}\right\vert =\protect\omega %
_{fe}-\protect\omega _{c_{l}}=\left\vert \protect\delta _{1}\right\vert
+\Delta _{1l},$ and $\Delta _{1l}=\protect\omega _{fg}-\protect\omega %
_{c_{1}}-\protect\omega _{c_{l}}>0$. Here, $\protect\omega _{c_{1}}$ ($%
\protect\omega _{c_{l}}$) is the frequency of cavity $1$ ($l$); $\protect%
\omega _{eg},$ $\protect\omega _{fe}$, and $\protect\omega _{fg}$ are the $%
|g\rangle \leftrightarrow |e\rangle ,$ $|e\rangle \leftrightarrow |f\rangle
, $ and $|g\rangle \leftrightarrow |f\rangle $ transition frequencies of the
qutrit, respectively. The red vertical line represents the frequency $%
\protect\omega _{c_{1}} $ of cavity $1,$ while the blue, green, ..., and
purple vertical lines represent the frequency $\protect\omega _{c_{2}}$ of
cavity $2$, the frequency $\protect\omega _{c_{3}}$ of cavity $3$,..., and
the frequency $\protect\omega _{c_{n}}$ of cavity $n$, respectively.}
\label{fig2}
\end{figure}

Consider $n$ microwave cavities ($1,2,...,n$) coupled to a
superconducting transmon qutrit [Fig.~\ref{fig1}(a)]. The three level of the
qutrit are denoted as $|g\rangle $, $|e\rangle $ and $|f\rangle $, as shown
in Fig.~\ref{fig1}(b). Theoretically, the $|g\rangle $ $\leftrightarrow $ $%
|f\rangle $ coupling for an ideal transmon is zero due to the selection rule
[50]. However, in practice, there exists a weak coupling between the two
levels $|g\rangle $ and $|f\rangle $ [51]. Suppose that cavity $1$ is
off-resonantly coupled to the $|g\rangle \leftrightarrow |e\rangle $
transition of the qutrit with coupling constant $g_{1}$ and detuning $%
\left\vert \delta _{1}\right\vert $ but highly detuned (decoupled) from the $%
|e\rangle \leftrightarrow |f\rangle $ transition of the qutrit. In addition,
assume that cavity $l$ ($l=2,3,...,n $) is off-resonantly coupled to the $%
|e\rangle \leftrightarrow |f\rangle $ transition of the qutrit with coupling
constant $g_{l}$ and detuning $\left\vert \delta _{l}\right\vert $ but
highly detuned (decoupled) from the $|g\rangle \leftrightarrow |e\rangle $
transition of the qutrit (Fig.~\ref{fig2}). Note that the coupling and
decoupling conditions can in principle be satisfied by prior adjustment of
the qutrit's level spacings or/and the cavity frequency. For a
superconducting (SC) qutrit, the level spacings can be rapidly (within 1-3
ns) adjusted by varying external control parameters (e.g., magnetic flux
applied to the superconducting loop of a SC phase, transmon [52], Xmon [53],
or flux qubit/qutrit [54]. In addition, the frequency of a microwave cavity
or resonator can be rapidly adjusted with a few nanoseconds [55,56].

Under the above assumptions, the Hamiltonian of the whole system, in the
interaction picture and after making the rotating-wave approximation (RWA),
can be written as (in units of $\hbar =1$)
\begin{equation}
H_{\mathrm{I}}=g_{1}(e^{-i\delta _{1}t}\hat{a}_{1}\sigma
_{eg}^{+}+h.c.)+\sum\limits_{l=2}^{n}g_{l}(e^{i\delta _{l}t}\hat{a}%
_{l}\sigma _{fe}^{+}+h.c.),
\end{equation}%
where $\sigma _{eg}^{+}=|e\rangle \langle g|$, $\sigma _{fe}^{+}=|f\rangle
\langle e|$, $\delta _{1}=\omega _{c_{1}}-\omega _{eg},$ and $\delta
_{l}=\omega _{fe}-\omega _{c_{l}}.$ To simplify Fig. 2, the detunings $%
\delta _{l}$ ($l=2,3,...,n$) are not drawn in Fig. \ref{fig2}, but their
definitions are given in the caption of Fig. \ref{fig2}. The detunings $%
\delta _{1}$ and $\delta _{l}$ have a relationship $\left\vert \delta
_{l}\right\vert =\left\vert \delta _{1}\right\vert +\Delta _{1l},$ with $%
\Delta _{1l}=\omega _{fg}-\omega _{c_{1}}-\omega _{c_{l}}>0$ (Fig. \ref{fig2}%
). Here, $\hat{a}_{1}$ ($\hat{a}_{l}$) is the photon annihilation operator
of cavity $1$ ($l$)$,$ $\omega _{c_{l}}$ is the frequency of cavity $l$ ($%
l=2,3,...,n$); while $\omega _{fe},$ $\omega _{eg},$ and $\omega _{fg}$ are
the $|e\rangle \leftrightarrow |f\rangle ,$ $|g\rangle \leftrightarrow
|e\rangle ,$ and $|g\rangle \leftrightarrow |f\rangle $ transition
frequencies of the qutrit, respectively.

Under the large-detuning conditions $\left\vert \delta _{1}\right\vert \gg
g_{1}$ and $\left\vert \delta _{l}\right\vert \gg g_{l}$, the Hamiltonian
(2) becomes [57]
\begin{eqnarray}
H_{\mathrm{e}}=& \lambda _{1}(\hat{a}_{1}^{+}\hat{a}_{1}|g\rangle \langle g|-%
\hat{a}_{1}\hat{a}_{1}^{+}|e\rangle \langle e|)  \nonumber \\
& -\sum\limits_{l=2}^{n}\lambda _{l}(\hat{a}_{l}^{+}\hat{a}_{l}|e\rangle
\langle e|-\hat{a}_{l}\hat{a}_{l}^{+}|f\rangle \langle f|)  \nonumber \\
& +\sum\limits_{l=2}^{n}\lambda _{1l}(e^{-i\bigtriangleup _{1l}t}\hat{a}%
_{1}^{+}\hat{a}_{l}^{+}\sigma _{fg}^{-}+h.c.)  \nonumber \\
& +\sum\limits_{k\neq l;k,l=2}^{n}\lambda _{kl}\left( e^{i\bigtriangleup
_{kl}t}\hat{a}_{k}^{+}\hat{a}_{l}+h.c.\right) \left( |f\rangle \langle
f|-|e\rangle \langle e|\right) ,
\end{eqnarray}%
where $\lambda _{1}=g_{1}^{2}/\left\vert \delta _{1}\right\vert $, $\lambda
_{l}=g_{l}^{2}/\left\vert \delta _{l}\right\vert $, $\lambda _{1l}=\left(
g_{1}g_{l}/2\right) (1/|\delta _{1}|+1/|\delta _{l}|)$, $\lambda
_{kl}=\left( g_{k}g_{l}/2\right) (1/|\delta _{k}|+1/|\delta _{l}|),$ $%
\bigtriangleup _{kl}=\left\vert \delta _{l}\right\vert -|\delta _{k}|$ $%
=\omega _{c_{k}}-\omega _{c_{l}},$ and $\sigma _{fg}^{-}=|g\rangle \langle
f| $. In Eq.~(3), the terms in the first two lines describe the photon
number dependent stark shifts of the energy levels $|g\rangle $, $|e\rangle $
and $|f\rangle $, the terms in the third line describe the $|f\rangle $ $%
\leftrightarrow $ $|g\rangle $ coupling caused due to the cooperation of
cavitie $1$ and $l$, while the terms in the last line describe the coupling
between cavities $k$ and $l.$ For $\bigtriangleup _{1l}\gg \{\lambda
_{1},\lambda _{l},\lambda _{1l}\}$, the effective Hamiltonian $H_{\mathrm{e}%
} $ changes to [57]
\begin{eqnarray}
H_{\mathrm{e}}=& \lambda _{1}(\hat{a}^{+}\hat{a}|g\rangle \langle g|-\hat{a}%
\hat{a}^{+}|e\rangle \langle e|)  \nonumber \\
& -\sum\limits_{l=2}^{n}\lambda _{l}(\hat{a}_{l}^{+}\hat{a}_{l}|e\rangle
\langle e|-\hat{a}_{l}\hat{a}_{l}^{+}|f\rangle \langle f|)  \nonumber \\
& +\sum\limits_{l=2}^{n}\chi _{1l}(\hat{a}_{1}\hat{a}_{1}^{+}\hat{a}_{l}\hat{%
a}_{l}^{+}|f\rangle \langle f|-\hat{a}_{1}^{+}\hat{a}_{1}\hat{a}_{l}^{+}\hat{%
a}_{l}|g\rangle \langle g|)  \nonumber \\
& +\sum\limits_{k\neq l;k,l=2}^{n}\lambda _{kl}\left( e^{i\bigtriangleup
_{kl}t}\hat{a}_{k}^{+}\hat{a}_{l}+h.c.\right) \left( |f\rangle \langle
f|-|e\rangle \langle e|\right) ,
\end{eqnarray}%
where $\chi _{1l}=\lambda _{1l}^{2}/\Delta _{1l}$. Eq.~(4) shows that each
term is associated with the level $|g\rangle $, $|e\rangle $, or $|f\rangle $%
. When the levels $|e\rangle $ and $|f\rangle $ are initially not occupied,
they will remain unpopulated under the Hamiltonian (4). This is because the
Hamiltonian (4) does not induce either $|g\rangle \rightarrow |e\rangle$
transition or $|g\rangle \rightarrow |f\rangle$ transition. Thus, the
effective Hamiltonian (4) reduces to
\begin{equation}
H_{\mathrm{e}}=\lambda _{1}\hat{n}_{1}|g\rangle \langle
g|-\sum\limits_{l=2}^{n}\chi _{1l}\hat{n}_{1}\hat{n}_{l}|g\rangle \langle g|,
\end{equation}%
where $\hat{n}_{1}=\hat{a}_{1}^{+}\hat{a}_{1}$ and $\hat{n}_{l}=\hat{a}%
_{l}^{+}\hat{a}_{l}$ are the photon number operators for cavities $1$ and $l$%
, respectively.

Suppose that the qutrit is initially in the ground state $\left\vert
g\right\rangle $. It will remain in this state throughout the interaction as
the Hamiltonian $H_{\mathrm{e}}$ cannot induce any transition for the
qutrit. In this case, the Hamiltonian $H_{\mathrm{e}}$ is reduced to
\begin{equation}
\widetilde{H}_{\mathrm{e}}=H_{0}+H_{\mathrm{int}},
\end{equation}%
with
\begin{eqnarray}
H_{0} &=&\lambda _{1}\hat{n}_{1},  \nonumber \\
H_{\mathrm{int}} &=&-\sum\limits_{l=2}^{n}\chi _{1l}\hat{n}_{1}\hat{n}_{l},
\end{eqnarray}%
which is the effective Hamiltonian governing the dynamics of the $n$
cavities ($1,2,...,n$).

Because of $[H_{0},H_{\mathrm{int}}]=0$, the unitary operator $U=e^{-i%
\widetilde{H}_{\mathrm{e}}t}$ can be written as
\begin{equation}
U=e^{-iH_{0}t}\otimes e^{-iH_{\mathrm{int}}t}=U_{1}\otimes
\prod\limits_{l=2}^{n}U_{1l},
\end{equation}%
where $U_{1}$ is a unitary operator on cavity $1,$ while $U_{1l}$ is a
unitary operator on cavities $1$ and $l,$ given by
\begin{eqnarray}
U_{1} &=&\exp \left( -i\lambda _{1}\hat{n}_{1}t\right) , \\
U_{1l} &=&\exp \left( i\chi _{1l}\hat{n}_{1}\hat{n}_{l}t\right) .
\end{eqnarray}

For a cqubit, the two logical states $|0\rangle $ and $|1\rangle $ are
encoded with cat states of a cavity, i.e.,
\begin{eqnarray}
|0\rangle  &=&N_{\alpha }^{+}(|\alpha \rangle +|-\alpha \rangle ),~
\nonumber \\
|1\rangle  &=&N_{\alpha }^{-}(|\alpha \rangle -|-\alpha \rangle ),
\end{eqnarray}%
where $N_{\alpha }^{\pm }=1/\sqrt{2(1\pm e^{-2|\alpha |^{2}})}$ are the
normalization coefficients. Because of $|\alpha \rangle =e^{-|\alpha
|^{2}/2}\sum\limits_{n=0}^{\infty }\frac{\alpha ^{n}}{\sqrt{n!}}|n\rangle $
and $|-\alpha \rangle =e^{-|\alpha |^{2}/2}\sum\limits_{n=0}^{\infty }\frac{%
(-\alpha )^{n}}{\sqrt{n!}}|n\rangle $, we have
\begin{eqnarray}
|0\rangle  &=&\sum\limits_{m=0}^{\infty }C_{2m}|2m\rangle ,\   \nonumber \\
\ |1\rangle  &=&\sum\limits_{n=0}^{\infty }C_{2n+1}|2n+1\rangle ,
\end{eqnarray}%
where $C_{2m}=2N_{\alpha }^{+}e^{-|\alpha |^{2}/2}\alpha ^{2m}/\sqrt{(2m)!}$
and $C_{2n+1}=2N_{\alpha }^{-}e^{-|\alpha |^{2}/2}\alpha ^{2n+1}/\sqrt{%
(2n+1)!}$. Eq.~(12) shows that the cat state $|0\rangle $ is orthogonal to
the cat state $|1\rangle $, independent of $\alpha $ (except for $\alpha =0$%
).

Based on Eq. (10) and Eq. (12), one can easily see that the unitary
operation $U_{1l}$ leads to the following state transformation
\begin{eqnarray}
U_{1l}|1_{1}1_{l}\rangle _{ab} &=&\sum\limits_{n,n^{\prime }=0}^{\infty
}\exp [i(2n+1)(2n^{\prime }+1)\chi _{1l}t]C_{2n+1}C_{2n^{\prime
}+1}|2n+1\rangle _{1}|2n^{\prime }+1\rangle _{l},  \nonumber \\
U_{1l}|1_{1}0_{l}\rangle  &=&\sum\limits_{n,m\prime =0}^{\infty }\exp
[i(2n+1)(2m^{\prime })\chi _{1l}t]C_{2n+1}C_{2m^{\prime }}|2n+1\rangle
_{1}|2m^{\prime }\rangle _{l},  \nonumber \\
U_{1l}|0_{1}0_{l}\rangle  &=&\sum\limits_{m,m^{\prime }=0}^{\infty }\exp
\left[ i(2m)(2m^{\prime })\chi _{1l}t\right] C_{2m}C_{2m^{\prime
}}|2m\rangle _{1}|2m^{\prime }\rangle _{l},  \nonumber \\
U_{1l}|0_{1}1_{l}\rangle  &=&\sum\limits_{m,n^{\prime }=0}^{\infty }\exp
[i(2m)(2n^{\prime }+1)\chi _{1l}t]C_{2m}C_{2n^{\prime }+1}|2m\rangle
_{1}|2n^{\prime }+1\rangle _{l}.
\end{eqnarray}%
For $\chi _{1l}t=\pi ,$ we have $\exp \left[ i(2m)(2m^{\prime })\chi _{1l}t%
\right] =\exp [i(2m)(2n^{\prime }+1)\chi _{1l}t]=\exp [i(2m)(2n^{\prime
}+1)\chi _{1l}t]=1$ but $\exp [i(2n+1)(2n^{\prime }+1)\chi _{1l}t]=-1.$
Thus, the state transformation (13) becomes
\begin{eqnarray}
U_{1l}|0_{1}0_{l}\rangle  &=&|0_{1}0_{l}\rangle ,  \nonumber \\
U_{1l}|0_{1}1_{l}\rangle  &=&|0_{1}1_{l}\rangle ,  \nonumber \\
U_{1l}|1_{1}0_{l}\rangle  &=&|1_{1}0_{l}\rangle ,  \nonumber \\
U_{1l}|1_{1}1_{l}\rangle  &=&-|1_{1}1_{l}\rangle ,
\end{eqnarray}%
which shows that the operator $U_{1l}$ implements a universal
controlled-phase gate on two cqubits $1$ and $l$, described by $%
|0_{1}0_{l}\rangle \rightarrow |0_{1}0_{l}\rangle $, $|0_{1}1_{l}\rangle
\rightarrow |0_{1}1_{l}\rangle $, $|1_{1}0_{l}\rangle \rightarrow
|1_{1}0_{l}\rangle $, and $|1_{1}1_{l}\rangle \rightarrow
-|1_{1}1_{l}\rangle $. It is obvious that the state transformation (14) can
be simplified as
\begin{eqnarray}
U_{1l}|0_{1}i_{l}\rangle |g\rangle  &=&|0_{1}i_{l}\rangle |g\rangle
\nonumber \\
U_{1l}|1_{1}i_{l}\rangle |g\rangle  &=&\left( -1\right)
^{i_{l}}|1_{1}i_{l}\rangle |g\rangle ,
\end{eqnarray}%
where $i_{l}$ $\in \left\{ 0,1\right\} .$

According to Eq. (15), it is easy to obtain the following state
transformation%
\begin{eqnarray}
\prod\limits_{l=2}^{n}U_{1l}\left\vert 0_{1}\right\rangle \left\vert
i_{2}\right\rangle \left\vert i_{3}\right\rangle ...\left\vert
i_{n}\right\rangle &=&\left\vert 0_{1}\right\rangle \left\vert
i_{2}\right\rangle \left\vert i_{3}\right\rangle ...\left\vert
i_{n}\right\rangle ,  \nonumber \\
\prod\limits_{l=2}^{n}U_{1l}\left\vert 1_{1}\right\rangle \left\vert
i_{2}\right\rangle \left\vert i_{3}\right\rangle ...\left\vert
i_{n}\right\rangle &=&\left\vert 1_{1}\right\rangle \left( -1\right)
^{i_{2}}\left( -1\right) ^{i_{3}}...\left( -1\right) ^{i_{n}}\left\vert
i_{2}\right\rangle \left\vert i_{3}\right\rangle ...\left\vert
i_{n}\right\rangle .
\end{eqnarray}

Now let us go back to the operator $U_{1}.$ According to (9) and (12), this
unitary operator leads to the following state transformation%
\begin{eqnarray}
U_{1}|0_{1}\rangle &=&\sum\limits_{m=0}^{\infty }\exp \left[ -i(2m)\lambda
_{1}t\right] C_{2m}|2m\rangle _{1},  \nonumber \\
U_{1}|1_{1}\rangle &=&\sum\limits_{n=0}^{\infty }\exp [-i(2n+1)\lambda
_{1}t]C_{2n+1}|2n+1\rangle _{1}.
\end{eqnarray}%
For $\lambda _{1}t=2\pi ,$ we have $\exp \left[ -i(2m)\lambda _{1}t\right]
=\exp [-i(2n+1)\lambda _{1}t]=1.$ Hence, Eq. (17) becomes
\begin{eqnarray}
U_{1}|0_{1}\rangle &=&|0_{1}\rangle ,  \nonumber \\
U_{1}|1_{1}\rangle &=&|1_{1}\rangle .
\end{eqnarray}%
Combining Eq. (16) and Eq. (18) leads to
\begin{eqnarray}
U_{1}\prod\limits_{l=2}^{n}U_{1l}\left\vert 1_{1}\right\rangle \left\vert
i_{2}\right\rangle \left\vert i_{3}\right\rangle ...\left\vert
i_{n}\right\rangle &=&\left\vert 1_{1}\right\rangle \left( -1\right)
^{i_{2}}\left( -1\right) ^{i_{3}}...\left( -1\right) ^{i_{n}}\left\vert
i_{2}\right\rangle \left\vert i_{3}\right\rangle ...\left\vert
i_{n}\right\rangle,  \nonumber \\
U_{1}\prod\limits_{l=2}^{n}U_{1l}\left\vert 0_{1}\right\rangle \left\vert
i_{2}\right\rangle \left\vert i_{3}\right\rangle ...\left\vert
i_{n}\right\rangle &=&\left\vert 0_{1}\right\rangle \left\vert
i_{2}\right\rangle \left\vert i_{3}\right\rangle ...\left\vert
i_{n}\right\rangle.
\end{eqnarray}%
The result (19) shows that when the control cqubit $1$ is in the state $%
\left\vert 0\right\rangle $, nothing happens to the states of each of target
cqubit ($2,3,...,n$); however, when the control cqubit $1\ $is in $%
\left\vert 1\right\rangle $, a phase flip (from sign $+$ to $-$) happens to
the state $\left\vert 1\right\rangle $ of each of target cqubits ($2,3,...,n$%
). Note that $U=U_{1}\otimes \prod\limits_{l=2}^{n}U_{1l}$ [see Eq. (8)].
Hence, a multi-target-qubit controlled phase gate, described by Eq. (1), is
implemented with $n$ cqubits ($1,2,...,n $), after the above operation
described by the unitary operator $U$.

From the description given above, one can see that the gate realization is
based on a single unitary operator $U$ which was obtained by starting with
the original Hamiltonian (2). Hence, the gate is implemented with a single
operation described by $U$. The qutrit remains in the ground state during
the gate operation. Therefore, decoherence from the qutrit is greatly
suppressed.

\begin{figure}[tbp]
\begin{center}
\includegraphics[bb=90 395 421 634, width=8.5 cm, clip]{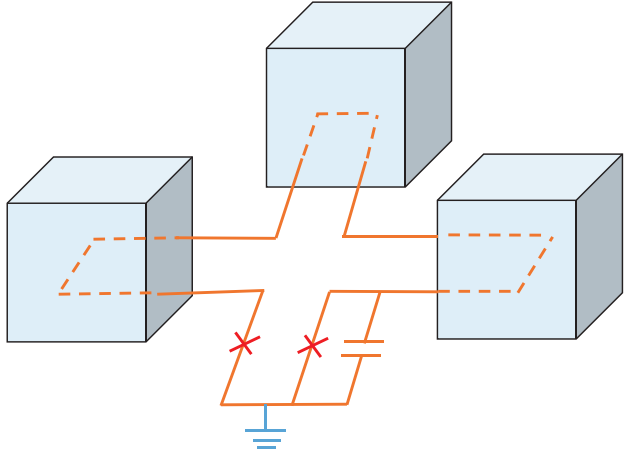} \vspace*{%
-0.08in}
\end{center}
\caption{(Color online) Setup for three 3D microwave cavities inductively
coupled to a transmon qutrit. Electronic circuit of a transmon qutrit
consists of two Josephson junctions and a capacitor.}
\label{fig3}
\end{figure}

As shown above, the conditions $\chi _{1l}t=\pi $ (independent $l$) and $%
\lambda _{1}t=2\pi $ should be met. They turn out into $\chi _{1l}=\lambda
_{1}/2$, which can be further written as%
\begin{equation}
g_{l}=\frac{\left\vert \delta _{l}\right\vert }{\left\vert \delta
_{1}\right\vert +\left\vert \delta _{l}\right\vert }\sqrt{2\Delta
_{1l}\left\vert \delta _{1}\right\vert }.
\end{equation}%
This condition (20) can be readily satisfied by varying $g_{l}$ or $%
\left\vert \delta _{l}\right\vert $ or both, given $\left\vert \delta
_{1}\right\vert $. Note that the detuning $\left\vert \delta _{l}\right\vert
$ can be adjusted by changing the frequency of cavity $l$, and the coupling
strength $g_{l}$ can be adjusted by a prior design of the sample with
appropriate capacitance or inductance between the qutrit and cavity $l$
[58,59].

Another point should be mentioned here. For circuit QED, the frequencies of
cavities ($2,3,...,n$) should be different in order to suppress the unwanted
inter-cavity crosstalk. Because of $\Delta _{1l}=\omega _{fg}-\omega
_{c_{1}}-\omega _{c_{l}}$ depending $\omega _{c_{l}},$ the detuning $\Delta
_{1l}$ would be different for cavities ($2,3,...,n$) with different
frequencies. However, for a cavity QED system consisting of cavities and a
natural atom (the coupler), there does not exist the inter-cavity crosstalk.
Thus, each of cavities ($2,3,...,n$) can be allowed to have the same
frequency, resulting in the same detuning $\Delta _{1l}$. This would
significantly reduce the experimental difficulty.

Before ending this section, we should point out that in quantum optics, the
two cat states described by Eq. (11) are called even and odd coherent
states, respectively. According to Ref.~[60], the encoding (11) for a
cat-state qubit works for a noise environment without phase damping
or a noise environment where phase damping is not dominant.

\section{Possible Experimental Implementation}

\label{sec-III}

For the sake of definitiveness, we give a brief discussion on the
experimental feasibility of implementing a three-qubit controlled phase gate
with one cqubit simultaneously controlling two target cqubits, by
considering a setup of a SC transmon qutrit coupled to three 3D microwave
cavities or resonators.

\begin{figure}[tbp]
\begin{center}
\includegraphics[bb=153 561 366 756, width=7.5 cm, clip]{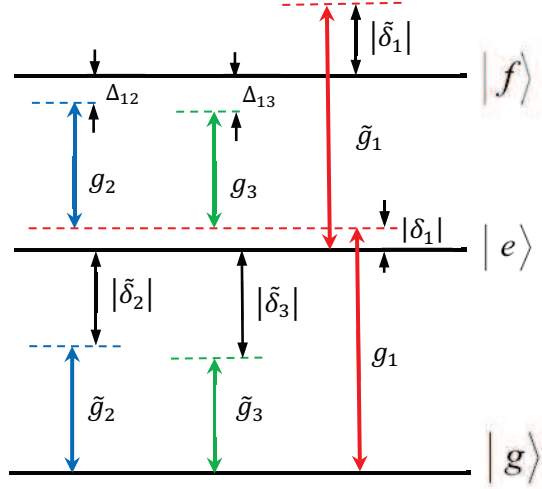} \vspace*{%
-0.08in}
\end{center}
\caption{(Color online) Illustration of the unwanted coupling between cavity
$1$ and the $|e\rangle \leftrightarrow |f\rangle $ transition of the qutrit
(with coupling strength $\widetilde{g}_{1}$ and detuning $\left\vert
\widetilde{\protect\delta }_{1}\right\vert $) as well as the unwanted
coupling between cavity $l$ and the $|g\rangle \leftrightarrow |e\rangle $
transition of the qutrit (with coupling strength $\widetilde{g}_{l}$ and
detuning $\left\vert \widetilde{\protect\delta }_{l}\right\vert $) ($l=2,3$%
). Here, $\left\vert \widetilde{\protect\delta }_{1}\right\vert =\protect%
\omega _{c_{1}}-\protect\omega _{fe}$ and $\left\vert \widetilde{\protect%
\delta }_{l}\right\vert =\protect\omega _{eg}-\protect\omega _{c_{l}}$ ($%
l=2,3$). Other couplings depicted in the figure are needed, as described by
the Hamiltonian (2) with $n=3$. Note that the coupling of each cavity with
the $|g\rangle \leftrightarrow |f\rangle $ transition of the qutrit is
negligible because of the forbidden or very weak $|g\rangle \leftrightarrow
|f\rangle $ transition [50,51].}
\label{fig4}
\end{figure}

From the description given above, one can see that the gate implementation
involves the operation, described by the Hamiltonian (2). In reality, the
inter-cavity crosstalk between cavities should be considered [61], and there
exist the unwanted coupling of cavity $1$ with the $|e\rangle
\leftrightarrow |f\rangle $ transition and the unwanted coupling of cavities
$2$ and $3$ with the $|g\rangle \leftrightarrow |e\rangle $ transition of
the qutrit (Fig.~\ref{fig4}). When these factors are taken into account, the
Hamiltonian (2) becomes%
\begin{equation}
\widetilde{H}_{\mathrm{I}}=H_{\mathrm{I}}+\Delta H+\varepsilon ,
\end{equation}%
with%
\begin{eqnarray}
\Delta H &=&\widetilde{g}_{1}(e^{-i\widetilde{\delta }_{1}t}\hat{a}%
_{1}\sigma _{fe}^{+}+h.c.)  \nonumber \\
&&+\sum\limits_{l=2}^{3}\widetilde{g}_{l}(e^{i\widetilde{\delta }_{l}t}\hat{a%
}_{l}\sigma _{eg}^{+}+h.c.),
\end{eqnarray}%
\begin{equation}
\varepsilon =\sum\limits_{k\neq l;k,l=1}^{3}g_{kl}(e^{-i\widetilde{\Delta }%
_{kl}t}\hat{a}_{k}\hat{a}_{l}^{+}+h.c.)
\end{equation}%
where $H_{\mathrm{I}}$ is the Hamiltonian (2) for $n=3$, $\Delta H$ is the
Hamiltonian describing the unwanted coupling between cavity $1$ and the $%
|e\rangle \leftrightarrow |f\rangle $ transition with coupling strength $%
\widetilde{g}_{1}$ and detuning $\widetilde{\delta }_{1}=\omega
_{c_{1}}-\omega _{fe}$ as well as the unwanted coupling between cavity $l$
and the $|g\rangle \leftrightarrow |e\rangle $ transition with coupling
strength $\widetilde{g}_{l}$ and detuning $\widetilde{\delta }_{l}=\omega
_{eg}-\omega _{c_{l}}$ ($l=2,3$) (Fig.~4); $\varepsilon $ represents the
inter-cavity crosstalk, where $g_{kl}$ is the coupling strength between
cavities $k$ and $l$ while $\widetilde{\bigtriangleup }_{kl}=\omega
_{c_{k}}-\omega _{c_{l}}$ is the difference between the frequencies of
cavities $k$ and $l$ ($k\neq l;k,l\in \left\{ 1,2,3\right\} $).

The dynamics of the lossy system is determined by
\begin{eqnarray}
\frac{d\rho }{dt}=& -i[\widetilde{H}_{\mathrm{I}},\rho
]+\sum_{l=1}^{3}\kappa _{l}\mathcal{L}[a_{l}]  \nonumber \\
& +\gamma _{eg}\mathcal{L}[\sigma _{eg}^{-}]+\gamma _{fe}\mathcal{L}[\sigma
_{fe}^{-}]+\gamma _{fg}\mathcal{L}[\sigma _{fg}^{-}]  \nonumber \\
& +\sum\limits_{j=e,f}\{\gamma _{\varphi j}(\sigma _{jj}\rho \sigma
_{jj}-\sigma _{jj}\rho /2-\rho \sigma _{jj}/2)\},
\end{eqnarray}%
where $\widetilde{H}_{\mathrm{I}}$ is the full Hamiltonian given above, $%
\sigma _{eg}^{-}=|g\rangle \langle e|$, $\sigma _{fe}^{-}=|e\rangle \langle
f|$, $\sigma _{fg}^{-}=|g\rangle \langle f|$, $\sigma _{jj}=|j\rangle
\langle j|(j=e,f)$; and $\mathcal{L}[\xi]=\xi\rho\xi^{+}-\xi^{+}\xi\rho/2-%
\rho\xi^{+}\xi/2$
with $\xi =a_{l},\sigma _{eg}^{-},\sigma _{fe}^{-},\sigma _{fg}^{-}$. In
addition, $\kappa _{l}$ is the photon decay rate of cavity $l$ $(l=1,2,3),$ $%
\gamma _{eg}$ is the energy relaxation rate for the level $|e\rangle $ of
the qutrit, $\gamma _{fe}(\gamma _{fg})$ is the energy relaxation rate of
the level $|f\rangle $ of the qutrit for the decay path $|f\rangle
\longrightarrow |e\rangle (|g\rangle )$, and $\gamma _{\varphi j}$ is the
dephasing rate of the level $|j\rangle (j=e,f)$ of the qutrit. \newline

The fidelity of the operation is given by
\begin{equation}
\mathcal{F}=\sqrt{\langle \psi _{\mathrm{id}}|\rho |\psi _{\mathrm{id}%
}\rangle },
\end{equation}%
where $|\psi _{\mathrm{id}}\rangle $ is the output state of an ideal system
without dissipation, dephasing and crosstalk; while $\rho $ is the final
practical density operator of the system when the operation is performed in
a realistic situation. The input state of the whole system is given by
\begin{eqnarray}
|\psi _{\mathrm{in}}\rangle &=&\left( c_{0}\left\vert 000\right\rangle
+c_{1}\left\vert 001\right\rangle +c_{2}\left\vert 010\right\rangle
+c_{3}\left\vert 011\right\rangle \right.  \nonumber \\
&&\left. +c_{4}\left\vert 100\right\rangle +c_{5}\left\vert 101\right\rangle
+c_{6}\left\vert 110\right\rangle +c_{7}\left\vert 111\right\rangle \right)
\otimes \left\vert g\right\rangle .
\end{eqnarray}%
where the coefficients $c_{0},c_{1},...,$and $c_{7}$ satisfy the
normalization condition $\sum_{k=0}^{7}\left\vert c_{k}\right\vert ^{2}=1.$
Thus, the ideal output state is
\begin{eqnarray}
|\psi _{\mathrm{id}}\rangle &=&\left( c_{0}\left\vert 000\right\rangle
+c_{1}\left\vert 001\right\rangle +c_{2}\left\vert 010\right\rangle
+c_{3}\left\vert 011\right\rangle \right.  \nonumber \\
&&\left. +c_{4}\left\vert 100\right\rangle -c_{5}\left\vert 101\right\rangle
-c_{6}\left\vert 110\right\rangle +c_{7}\left\vert 111\right\rangle \right)
\otimes \left\vert g\right\rangle .
\end{eqnarray}%
For simplicity, we choose
\begin{eqnarray}
c_{0} &=&\cos \gamma \cos \theta \cos \varphi ;~ c_{4}=\sin \gamma \cos
\theta \cos \varphi ,  \nonumber \\
c_{1} &=&\cos \gamma \cos \theta \sin \varphi ;~ c_{5}=\sin \gamma \cos
\theta \sin \varphi ,  \nonumber \\
c_{2} &=&\cos \gamma \sin \theta \cos \varphi ;~ c_{6}=\sin \gamma \sin
\theta \cos \varphi ,  \nonumber \\
c_{3} &=&\cos \gamma \sin \theta \sin \varphi ;~ c_{7}=\sin \gamma \sin
\theta \sin \varphi .
\end{eqnarray}%
In the following, we will consider the cases: (a) $\gamma =\theta =\varphi
=\pi /4;$ (b) $\gamma =\theta =\varphi =\pi /3;$ (c) $\gamma =\pi /2,\theta
=\pi /4,\varphi =\pi /3;$ and (d) $\gamma =\pi ,\theta =\pi /3,\varphi =\pi
/4;$ which correspond to four initial states.

\begin{figure}[tbp]
\begin{center}
\includegraphics[bb=0 0 346 265, width=12.5 cm, clip]{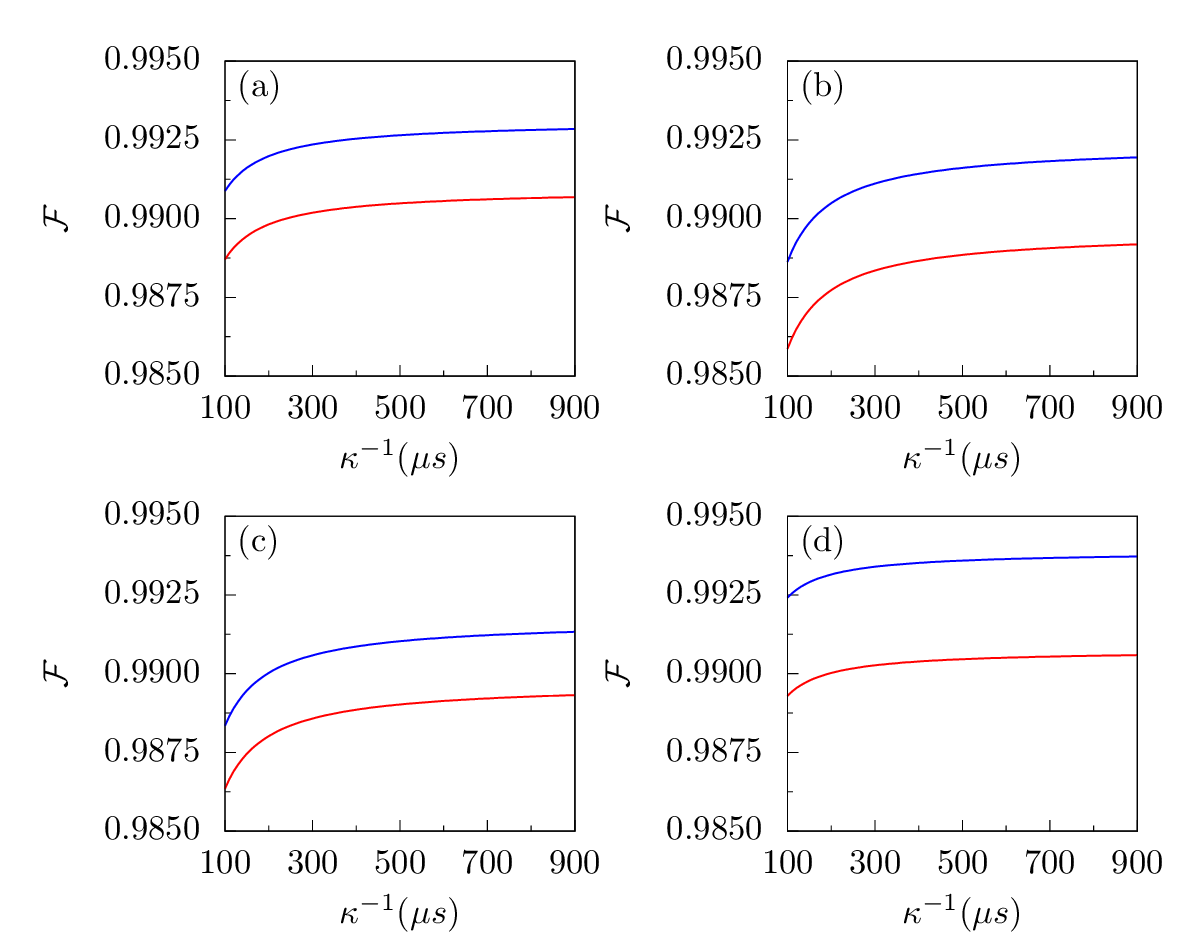} \vspace*{%
-0.08in}
\end{center}
\caption{(Color online) Fidelity versus $\protect\kappa ^{-1}$. The plots
are drawn for $\protect\alpha =0.5$. Other parameters used in the numerical
simulation are referred to the text. Blue curves are based on the effective
Hamitonian (5) and considering decoherence and the inter-cavity crosstalk;
while red curves are based on the full Hamiltonian (21) and considering
decoherence and the inter-cavity crosstalk. (a) is plotted for $\protect%
\gamma =\protect\theta =\protect\varphi =\protect\pi /4;$ (b) is for $%
\protect\gamma =\protect\theta =\protect\varphi =\protect\pi /3;$ (c) is for
$\protect\gamma =\protect\pi/2,\protect\theta =\protect\pi /4,\protect%
\varphi =\protect\pi /3$; while (d) is for $\protect\gamma =\protect\pi ,%
\protect\theta =\protect\pi/3,\protect\varphi =\protect\pi /4$.}
\label{fig:4}
\end{figure}

For a transmon qutrit, the typical transition frequency between neighboring
levels can be made as 3 to 15 GHz and the anharmonicity $100\sim 500$ MHz of
the level spacings has been reported in experiments [62,63]. As an example,
we thus consider $\omega _{eg}/2\pi =6.5$ GHz and $\omega _{fe}/2\pi =6.2$
GHz. By choosing $\left\vert \delta _{1}\right\vert /2\pi =0.5$ GHz, $%
\left\vert \delta _{2}\right\vert /2\pi =0.51$ GHz, and $\left\vert \delta
_{3}\right\vert /2\pi =0.52$ GHz, we have $\Delta _{12}/2\pi =0.01$ GHz, $%
\Delta _{13}/2\pi =0.02$ GHz, $\omega _{c_{1}}/2\pi =7.0$ GHz, $\omega
_{c_{2}}/2\pi =5.69$ GHz, and $\omega _{c_{3}}/2\pi =5.68$ GHz, for which we
have $\widetilde{\triangle }_{12}/2\pi =1.31$ GHz, $\widetilde{\triangle }%
_{23}/2\pi =0.01$ GHz, and $\widetilde{\triangle }_{13}/2\pi =1.32$ GHz.
With the transition frequencies of the qutrit and the frequecies of the
cavities given here, we have $\left\vert \widetilde{\delta }_{1}\right\vert
/2\pi =0.8$ GHz, $\left\vert \widetilde{\delta }_{2}\right\vert /2\pi =0.81$
GHz, and $\left\vert \widetilde{\delta }_{3}\right\vert /2\pi =0.82$ GHz.
Other parameters used in the numerical simulation are: (i) $\gamma
_{eg}^{-1}=60$ $\mu $s, $\gamma _{fg}^{-1}=150$ $\mu $s [64], $\gamma
_{fe}^{-1}=30$ $\mu $s, $\gamma _{\phi e}^{-1}=\gamma _{\phi f}^{-1}=20$ $%
\mu $s, (ii) $g_{1}/2\pi =35$ MHz, and (iii) $\alpha =0.5$. Here, we
consider a rather conservative case for decoherence time of the transmon
qutrit because energy relaxation time with a range from 65 $\mu $s to 0.1 ms
and dephasing time from 25 $\mu $s to 70 $\mu $s have been experimentally
reported for a 3D superconducting transmon device [33,65,66]. According to
Eq. (20), one can calculate the $g_{2}$ and $g_{3}$, which are $g_{2}/2\pi
\sim 50.5$ MHz and $g_{3}/2\pi \sim 72.1$ MHz. For a transmon qutrit [50],
one has $\widetilde{g}_{1}/2\pi \sim \sqrt{2}g_{1}/2\pi \sim 49.5$ MHz$,$ $%
\widetilde{g}_{2}/2\pi \sim g_{2}/(2\pi \sqrt{2})\sim 35.7$ MHz$,$ and $%
\widetilde{g}_{3}/2\pi \sim g_{3}/(2\pi \sqrt{2})\sim 41.6$ MHz. Note that
the coupling constants chosen here are readily available because a coupling
constant $\sim 2\pi \times 360$ MHz has been reported for a transmon device
coupled to a microwave cavity [67]. We set $g_{kl}=0.01g_{\max }$, where $%
g_{\max }=\max \{g_{1},g_{2},g_{3}\}\sim 2\pi \times 72.1$ MHz, which can be
achieved in experiments [33]. In addition, assume $\kappa _{1}=\kappa
_{2}=\kappa _{3}=\kappa $ for simplicity.

By solving the master equation (24), we numerically calculate the fidelity
versus $\kappa ^{-1}$, as depicted in Fig.~5. Fig. 5(a) is plotted for $%
\gamma =\theta =\varphi =\pi /4.$ Fig. 5(b) is for $\gamma =\theta =\varphi
=\pi /3.$ Fig. 5(c) is for $\gamma =\pi /2,\theta =\pi /4,\varphi =\pi /3.$
Fig. 5(d) is for $\gamma =\pi ,\theta =\pi /3,\varphi =\pi /4$. In Fig. 5,
the red curves are plotted by numerical simulations, based on the full
Hamiltonians $\widetilde{H}_{\mathrm{I}}$ and by taking decoherence and the
inter-cavity crosstalk into consideration. From the red curves, one can see
that when $\kappa ^{-1}\geq 300$ $\mu $s, fidelity exceeds: (i) $0.9902$ for
$\gamma =\theta =\varphi =\pi /4;$ (ii) $0.9884$ for $\gamma =\theta
=\varphi =\pi /3;$ (iii) $0.9886$ for $\gamma =\pi /2,\theta =\pi /4,\varphi
=\pi /3;$ and (iv) $0.9903$ for $\gamma =\pi ,\theta =\pi /3,\varphi =\pi /4.
$ These results imply that the fidelity varies with different initial states
of the three cavities, but a high fidelity can be obtained for the gate
being performed in a realistic situation.

We also calculate the fidelity, based on the effective Hamiltonian $H_{%
\mathrm{e}}$ in Eq.~(5) and by considering decoherence and the inter-cavity
crosstalk (see the blue curves in Fig. 5). From the red curves and the bule
curves in Fig. 5, it can be concluded that the fidelity for the gate
performed in a realistic situation is slightly decreased by $0.2\%-0.5\%,$
when compared to the case of the gate performed based on the effective
Hamiltonian. This result implies that the approximations, which we made
above for the effective Hamiltonian, are reasonable.

The gate operational time is estimated as $\sim 0.41$ $\mu $s for the
parameters chosen above, which is much shorter than the decoherence times of
the qutrit used in the numerical simulation and the cavity decay times ($100$
$\mu $s $-$ $900$ $\mu $s) considered in Fig. 5. Note that lifetime $\sim 1$
ms of microwave photons has been experimentally demonstrated in a 3D
resonator [46,68]. For the cavity frequencies given above and $\kappa
^{-1}=300$ $\mu $s, one has $Q_{1}\sim 1.31\times 10^{7}$ for cavity $1$, $%
Q_{2}\sim 1.07\times 10^{7}$ for cavity $2$, and $Q_{3}\sim 1.07\times 10^{7}
$ for cavity $3$, which are available because a high quality factor $%
Q=3.5\times 10^{7}$ of a 3D superconducting cavity has been experimentally
reported [68]. The analysis here implies that high-fidelity realization of a
quantum controlled phase gate with one cqubit simultaneously controlling two
target cqubit is feasible with the present circuit QED technology.

\section{Conclusions}

\label{sec-con}

We have proposed a one-step method to realize an $n$-qubit controlled phase
gate with one cat-state qubit simultaneously controlling $n-1$ target
cat-state qubits, based on circuit QED. This method can be applied to
implement the proposed gate with a wide range of physical systems, such as
multiple microwave or optical cavities coupled to a single three-level
natural or artificial atom. As shown above, this proposal has the following
features: (i) During the gate operation, the qutrit remains in the ground
state; thus decoherence from the qutrit is greatly suppressed; (ii) Because
only one-step operation is needed and neither classical pulse nor
measurement is required, the gate realization is simple; and (iii) The gate
operation time is independent of the number of the cat-state qubits. Our
numerical simulations show that high-fidelity implementation of a
three-qubit controlled phase gate with one cat-state qubit simultaneously
controlling two target cat-state qubits is feasible with current circuit QED
technology. To the best of our knowledge, this work is the first to
demonstrate the implementation of a multi-target-qubit controlled phase gate
with cat-state qubits based on cavity- or circuit-QED. We hope that this
work will stimulate experimental activities in the near future.

\section*{Acknowledgments}

This work was supported in part by the NKRDP of China (Grant No.
2016YFA0301802) and the National Natural Science Foundation of China under
Grant Nos. [11074062, 11374083,11774076]. This work was also supported by
the Hangzhou-City grant for Quantum Information and Quantum Optics
Innovation Research Team.


\begin{thebibliography}{99}
\bibitem{s64} For a transmon qutrit, the $|g\rangle $ $\leftrightarrow $ $%
|f\rangle $ transition is much weaker than those of the $|g\rangle $ $%
\leftrightarrow $ $|e\rangle $ and $|e\rangle $ $\leftrightarrow $ $%
|f\rangle $ transitions. Thus, we have $\gamma _{fg}^{-1}\gg \gamma
_{eg}^{-1},\gamma _{fe}^{-1}$.
\end{thebibliography}
\end{document}